\documentclass[12pt]{elsarticle}

\usepackage{amssymb}
\usepackage{amsthm}
\usepackage{dirtytalk}

\usepackage[left=2.5cm, right=2.5cm, top=4cm, bottom=3cm, footskip=0.5cm]{geometry}
\usepackage{graphicx}
\usepackage{subcaption}

\journal{ }

\begin{document}

\begin{frontmatter}

\title{Evaluating Supply Chain Resilience During Pandemic Using Agent-based Simulation}
\date{ }

\author[1,2,*]{Teddy Lazebnik}

\address[1]{Department of Mathematics, Ariel University, Ariel, Israel}
\address[2]{Department of Cancer Biology, Cancer Institute, University College London, London, UK}
\address[*]{Corresponding author: lazebnik.teddy@gmail.com}

\begin{abstract}
Recent pandemics have highlighted vulnerabilities in our global economic systems, especially supply chains. Possible future pandemic raise a dilemma for businesses owners between short-term profitability and long-term supply chain resilience planning. In this study, we propose a novel agent-based simulation model integrating extended Susceptible-Infected-Recovered (SIR) epidemiological model and supply and demand economic model to evaluate supply chain resilience strategies during pandemics. Using this model, we explore a range of supply chain resilience strategies under pandemic scenarios using \textit{in silico} experiments. We find that a balanced approach to supply chain resilience performs better in both pandemic and non-pandemic times compared to extreme strategies, highlighting the importance of preparedness in the form of a better supply chain resilience. However, our analysis shows that the exact supply chain resilience strategy is hard to obtain for each firm and is relatively sensitive to the exact profile of the pandemic and economic state at the beginning of the pandemic. As such, we used a machine learning model that uses the agent-based simulation to estimate a near-optimal supply chain resilience strategy for a firm. The proposed model offers insights for policymakers and businesses to enhance supply chain resilience in the face of future pandemics, contributing to understanding the trade-offs between short-term gains and long-term sustainability in supply chain management before and during pandemics.

\end{abstract}


\end{frontmatter}

\section{Introduction}
\label{sec:intro}
Pandemics, such as the recent global COVID-19 crisis \cite{covid_2,covid_1} or more historical ones such as the Spanish Influenza during World War I \cite{spanish_1918_ww1,spanish_1918_history}, have starkly illuminated the vulnerabilities in our economic systems \cite{covid_eco_1,covid_eco_2}. In the short term, pandemics disrupt production, significantly shift consumer demand, and strain healthcare resources, leading to immediate economic downturns \cite{economic_crisis_covid}. The long-term effects are equally concerning, with industries facing restructuring \cite{adam2020after,mckibbin2020global1}, labor market shifts \cite{teddy_paper2,del2020supply}, and altered consumer behavior patterns \cite{consumer_behavior}, all of which pose substantial challenges to economic recovery and growth. 

In particular, supply chains, as the lifelines of global commerce, are especially sensitive to the shocks induced by pandemics \cite{sc_1,sc_new}. The intricate web of interconnected suppliers, manufacturers, distributors, and retailers can quickly unravel under the strain of widespread disruptions, leading to shortages, price volatility, and logistical bottlenecks \cite{intro_sc_claim_1,intro_sc_claim_2}. The fragility of these supply chains has been laid bare during recent crises, prompting a critical reevaluation of resilience strategies \cite{sc_resilence_general}.

While the inevitability of pandemics is widely acknowledged \cite{pandemic_important,history_pandemic_changes}, businesses face a dilemma in balancing the need to prepare for future disruptions against the imperative to generate profits in the present \cite{intro_exp_1}. This \say{greedy} mindset, driven by short-term financial goals, often conflicts with the long-term resilience planning required to withstand future shocks \cite{future_prepare_economy}. This dilemma is similar to other dual-objective optimization tasks with conflicted agendas such as food exploration by ants \cite{ants} to more complex ones such as investment portfolio optimization tasks \cite{investment_protfolio}. Generally speaking, this tension between immediate profitability and future preparedness underscores the complexity of decision-making in a volatile and uncertain environment \cite{ouu_review,ouu_review_2}.

Addressing this tension presents an intriguing computational challenge \cite{sc_2,sc_3,sc_4}. How can businesses optimize their supply chain strategies to simultaneously maximize current profits and enhance resilience against future pandemics? Previous studies tried to address this question by analyzing strategies applied by businesses during previous pandemics and analyzing which businesses better handled the crisis \cite{sc_r_1,sc_r_2}. These studies indeed provide a fruitful ground for policy-makers and business owners to design their strategy but actually susceptible to Lucas's critique as well \citep{lucas_critique}. Namely, one would require a set of businesses to try two or more strategies to see which one works best for multiple pandemic configurations to be able to empirically establish a claim. Unfortunately, such an experiment is impossible in practice. 

To this end, mathematical models and computer simulation emerge as powerful tools to overcome this challenge \cite{math_model_1,math_model_2,math_model_3}. While limited in their expressiveness and accuracy in predicting the real world, they commonly provide accurate enough predictions to establish a reasonable replica (commonly referred to as \say{digital twin}) of the studied case \cite{digital_twin}. Specifically, agent-based simulation (ABS) is a computational method to describe the dynamics that occur due to the interaction of diverse agents \cite{abs_first_cite}. In this case, the agents represent different supply chain entities and the dynamic interactions between them \cite{abs_2,abs_3}. By simulating various scenarios and resilience strategies, one can gain valuable insights into effective approaches for navigating the delicate balance between short-term gains and long-term sustainability in the face of pandemics \cite{abs_1,abs_4}.

Indeed, previous studies investigate supply chain resilience during a crisis, in general, and for cases of a pandemic, in particular \cite{general_ref_1,general_ref_2,general_ref_3}. For instance, \cite{intro_exp_3} adopted the ABS method to analyze the influence of shifts in supply and demand due to the COVID-19 pandemic on the supply chain's ability to deliver. The authors studied two main recovery strategies relevant to building emergency supply and extra manufacturing capacity to mitigate supply chain disruptions using their model. \cite{basepaper} suggested analyzing and designing supply chain reliance from an immune system perspective as immune systems are safeguarding against disruptions and facilitating recovery - properties also associated with a resilient supply chain. The author proposed a mathematical formalization for supply chain resilience based on the biologically-inspired framework of immune systems. In a more general sense, \cite{intro_exp_2} reviewed supply chain resilience work with a focus on connecting the supply chain to other networks such as command-and-control and transportation. The author identified three main modeling strategies - linear, branching, and graph-based and concluded that graph-based models provide the most realistic and accurate modeling strategy of these. 

In this study, we present a novel ABS-based model to explore how much a business should focus its resources on supply chain resilience during a pandemic. The novelty of this work lies in the integration of a well-established pandemic-spread model based on an extended SIR (Susceptible-Infectious-Recovered) epidemiological model with theoretically and empirically proven economic supply-chain model which emerges from supply-and-demand dynamics and their shifts due to the pandemic. To be exact, the proposed model extends previous similar models by investigating pre-pandemic resource allocation towards supply chain resilience, which, as far as we know, has been mostly neglected. We first show how pandemics influence firms for different supply chain resilience strategies. Then, we explore the sensitivity of these strategies to economic and epidemiological changes, and finally, we use machine learning to estimate the supply chain resilience strategy a firm should adopt.  

The rest of this paper is organized as follows. Section \ref{sec:related_work} provides a quick overview of how supply chains are designed and utilized. In addition, an overview of the ABS method is presented alongside its usage for both epidemiological and supply chain models. Section \ref{sec:methods} presents the formalization of the proposed ABS simulation to explore supply chain resilience and experimental setup. Section \ref{sec:results} outlines the obtained results from the experiments. Finally, Section \ref{sec:conclusion} discusses the results in an economic context with its possible application and suggests possible future work. 

\section{Related Work}
\label{sec:related_work}
In this section, we provide an overview of how supply chains are designed, established, and managed. These properties are later taken into consideration in the modeling phase. In addition, an overview of the ABS method is presented in general and in the context of both epidemiological and supply chain models. These models are later used as the modeling fundamentals for our model.

\subsection{Supply chains}
Supply chains play a pivotal role in modern economies, encompassing the design, establishment, and management of interconnected networks that facilitate the flow of goods and services \cite{sc_rw_g_1,sc_rw_g_2}. Indeed, in the modern economy, individual companies no longer compete solely based on their unique brand identities. Instead, they operate as interconnected parts of supply chains. Success now hinges on a firm's ability to manage and coordinate the complex network of relationships within these supply chains \cite{sc_rw_p}. A supply chain functions as an integrated system, coordinating processes from sourcing raw materials to delivering finished products, adding value, promoting and distributing them, and facilitating information exchange among various entities like suppliers, manufacturers, distributors, logistics providers, and retailers. The primary goal is to boost operational efficiency, profitability, and competitive advantage for both the firm and its supply chain partners \cite{sc_rw_p}. Supply chain management is succinctly defined as the integration of crucial business processes, spanning from end-users to original suppliers, which adds value for customers and stakeholders \cite{sc_rw_p}.

Multiple approaches have been suggested for modeling supply chains. According to \cite{sc_rw_p_2_source}, these can be categorized into four groups: deterministic models with known parameters, stochastic models with at least one unknown parameter following a probabilistic distribution, economic game-theoretic models, and simulation-based models assessing supply chain strategy performance. Most of these models are steady-state, focusing on average performance or steady conditions \cite{sc_rw_p_2}. However, static models fall short in capturing the dynamic nature of supply chain systems affected by demand variations, lead-time delays, sales forecasting, to name a few \cite{sc_rw_p_2}. Multiple works show that a combination of supply and demand model with network analysis is appropriate to capture the complexity of supply chain management \cite{sc_rw_end_1}. This mathematical framework is often solved using ABS \cite{sc_rw_end_2}.

\subsection{Agent based simulation}
Agent based simulation (ABS) is a computational method of capturing (spatio-)temporal dynamics occurring for multiple agents \cite{abs_g_1,abs_g_2}. An ABS more often than not contains two main components - an environment and a population of agents which can be homogeneous or heterogeneous \cite{abs_h_1,abs_h_2}. ABS is based on three types of interactions between the agents and their environment - spontaneous, agent-agent, and agent-environment. Spontaneous interactions are interactions between an agent and itself that only depend on the agent's current state and time. Agent-agent interactions are interactions between two or more agents that alter the state of at least one of the agents taking part in the interaction. Agent-environment interactions are interactions between agents and their environment that change either or both the agent's state or the environment's state. Interestingly. ABS can be computationally reduced to the population protocol model \cite{population_protocol} and therefore it is Turing-complete \cite{turing_1,turing_2} meaning the ABS can express any dynamics solvable by a computer. 

A growing body of work utilizes the ABS computational method for a wide range of tasks \cite{abs_useful}. In particular, we below examine the usage of ABS in epidemiological and supply chain models. 

\subsubsection{Epidemiological models}
ABS is commonly used in the context of epidemiological models to model and solve heterogeneous population dynamics as other methods such as differential equations or functional models are limited in their ability to efficiently describe such dynamics \cite{abs_epi_g_1,abs_epi_g_2}. For example, \cite{abs_epi_exp_1} proposed an ABS of pedestrian dynamics to evaluate the behavior of pedestrians in public places in the context of contact transmission of airborne infectious diseases. The authors used a continuous space with a random direct walk of agents and infectious dynamics as a function of distance between the agents. \cite{abs_epi_exp_2} described a ABS based model of pandemic spread in facilities which is based on the popular SIR epidemiological model \cite{first_sir_paper} that assumes three epidemiological states - susceptible, infected, and recovered. In their model, agents had heterogeneous movement dynamics which were accomplished by three rules that take into consideration the agent's epidemiological state, as well as its local environment and the agent's in this environment. \cite{lazebnik2023high} proposed an ABS based model with an SEI (susceptible-exposed-infected) epidemiological model operating in a single room with three-dimensional geometry. The authors combined airflow dynamics using the Computational Fluid Dynamics (CFD) model with the epidemiological dynamics for airborne pathogens where agents are spatially static but have heterogeneous breathing patterns. 

\subsubsection{Supply chain models}
Similar to the ABS usage in epidemiological models, ABS is used for supply chain models to capture \say{economic players} which different objectives, capabilities, or roles in an economy, in general, and in a supply chain, in particular \cite{abs_sc_g,abs_sc_g_2}. For instance, \cite{abs_sc_exp_1} proposed an ABS based model of four three-level supply chains that apply different types of combined contracts by taking into account the effects of vertical and horizontal competition between supply chains. The authors show that the simulated results agree with previous socio-economic knowledge from the literature. \cite{abs_sc_exp_2} introduced an integrated framework for ABS inventory–production–transportation modeling and distributed simulation of supply chains. The authors show that their framework produces predictions that agree with previous known dynamics such as utilization of machines in the manufacturer and quantity change of products. \cite{abs_sc_exp_3} investigate retail stockouts using an ABS based model. The authors consider the change in market share as a measure of resilience for both the manufacturer and the retailer to examine the impact of the stockout and using the model, explore the effect of different scenarios on these metrics. 

\section{Methods and materials}
\label{sec:methods}
In this section, we initially introduce the proposed epidemiological-economic model for supply chain resilience management based on the ABS method. Afterward, we describe a supply chain resilience formalization with strategies firms can adopt. Next, we outline a machine learning strategy to obtain an approximation to the right balance of supply chain resilience and profit using a machine learning algorithm. Finally, an experimental setup for the model is outlined.

\subsection{Model definition}
The proposed epidemiological-economic model for supply chain resilience management is based on the ABS approach and constructed from three interconnected sub-models: epidemiological, economic, and supply chain (spatial). These sub-models are structured on top of three types of agents - consumers (which also function as workers), firms, and products. 

\begin{figure}[!ht]
    \centering
    \includegraphics[width=0.99\textwidth]{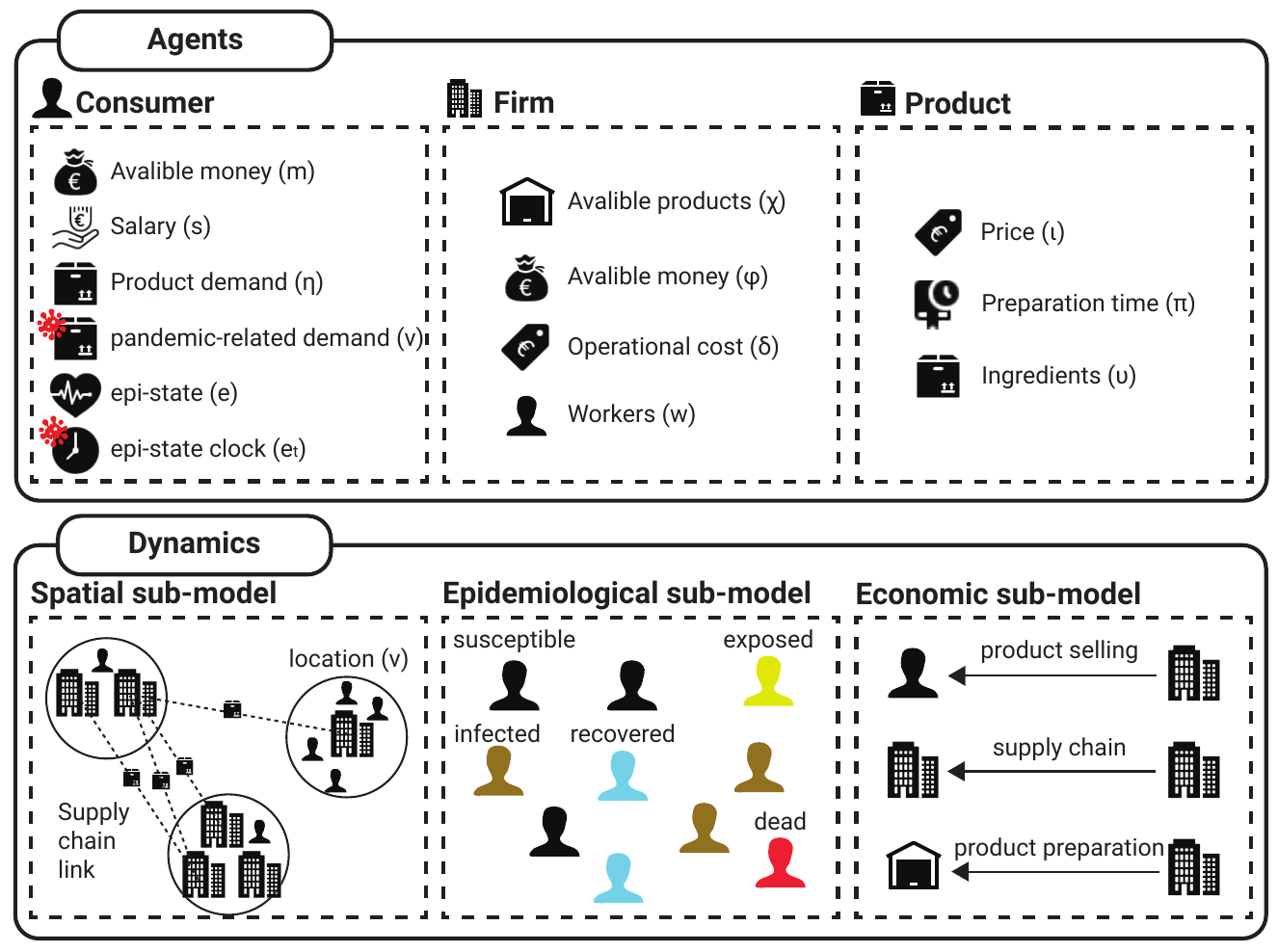}
    \caption{A schematic view of the proposed epidemiological-economic model with supply chains.}
    \label{fig:model_scheme}
\end{figure}

We define the model as a tuple \(\mathbb{M} := (C, F, P, G)\), where \(C\) is a set of consumers, \(F\) is a set of firms, \(P\) is a set of products, \(G\) is a graph of locations that the consumers, firms, and products are physically located in and interacting with each other and between themselves. The components of the tuple are described below in detail. Fig. \ref{fig:model_scheme} provides a schematic view of the model’s components and the interactions between them. 

Following the ABS method, we first formally define the three types of agents. In our model, all agents are represented by a timed finite state machine \cite{fsm}. The consumer agent, \(c \in C\) is defined by the tuple \(c := (m, s, \eta, \nu, e, e_t)\) where \(m \in [0, \infty)\) is the currently available money of the consumer, \(s \in [0, \infty)\) is the amount of salary the consumer gets in each step in time, \(\eta \in \mathbb{N}^{|P|}\) is the demand for products, \(\nu \in \mathbb{R}^{|P|}\) is the change for products' demand due to the pandemic state, \(e \in \{S, E, I, R, D\}\) is the consumer's epidemiological state, and \(e_t \in \mathbb{N}\) is the number of steps in time passed since the last time the consumer's epidemiological state changed. A firm \(f \in F\) is defined by the tuple \(f := (\chi, \varphi, \delta, w)\) where \(\chi \in \mathbb{N}^{|P|}\) is the number of currently available products to supply, \(\varphi \in [0, \infty)\) is the currently available money of the firm, \(\delta \in \mathbb{R}^+\) is the operational cost of the firm, \(w \subset C\) is the set of consumers that are also workers of the firm. A product \(p \in P\) is defined by the tuple \(p := (\iota, \pi, \upsilon)\) where \(\iota  \in [0, \infty) \) is the price of the product, \(\pi \in [0, \infty)\) is the preparation time of the product from its ingredients, and \(\upsilon \in \mathbb{N}^{|P|}\) is the vector of ingredients required to prepare a product. 

\begin{figure}[!ht]
    \centering
    \includegraphics[width=0.5\textwidth]{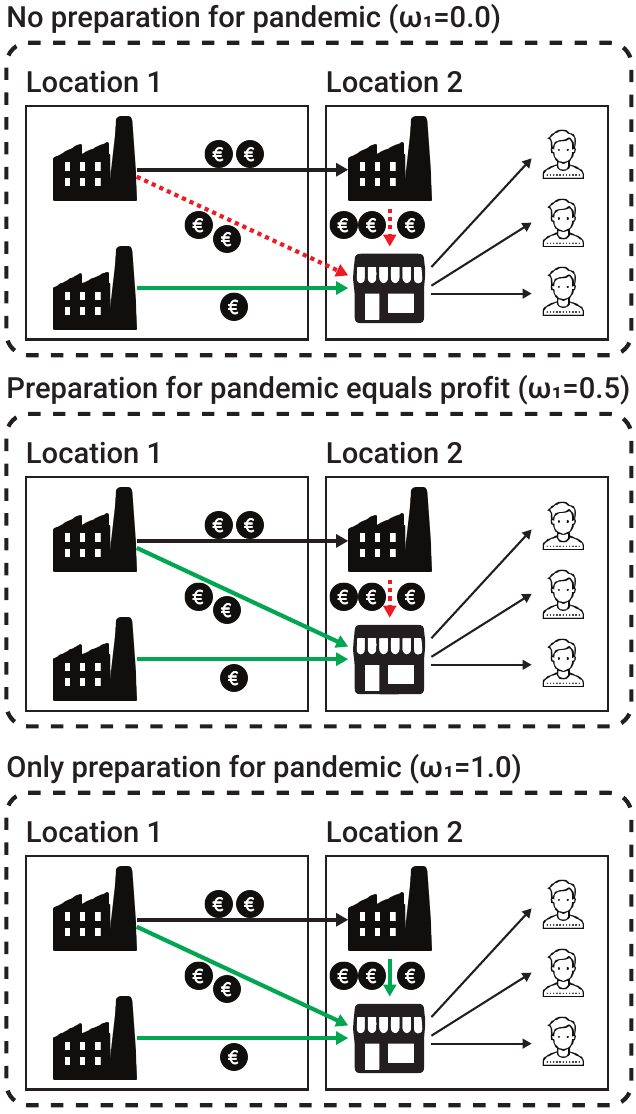}
    \caption{A simple example of three supply chain strategies obtained for different objectives of a firm from the perspective of a store (firm) for a case of only two locations and four firms where the consumer population is present only in the same location as the store. The solid-green arrows indicate the store firm established these supply chains while dashed-red arrows indicate the store firm do not established these supply chains.}
    \label{fig:example}
\end{figure}

The interaction between these agents are the core of the model and is captured by the three sub-models. Since there are many moving parts in the model, let us consider a simple example to capture the underlying behavior of the model. Let us consider two locations such that the first one has two firms and no consumer population while the other has also two firms and some consumer population. In this scenario, three out of the four firms will be factories as they produce products that do not have any ingredients and sell them to the fourth firm which operates as a store. As each firm (i.e. factory)  has different processes, the price of its product is different. For this example, let us assume each of the firms operating as factories are able to supply all the demand the firm operating as a store has in times of no pandemic. Here, we focus on the firm which operates as a store. If this firm aims to make as much profit as possible and ignores the supply chain realisance, it should buy the product from the firm that offers it for the cheapest price. On the other extreme case where this firm is only worried by the pandemic, establishing all possible supply chains will ensure the ability to satisfy demand. A balanced objective would cause a more diverse supply chain strategy while also considering profits. Fig. \ref{fig:example} presents a schematic representation of this example.  

\subsubsection{Epidemiological model}
For each location, the epidemiological sub-model is an extension of the SIR model and considers a constant consumer population with a fixed number of consumers (\(C_i\)) of size \(N_i\) for the \(i_{th}\) location. In the context of studying the immediate economic influence of the pandemic on the supply chain, the time horizon of interest is relatively short, ranging for up to several months, and as such, population growth can be neglected. Each consumer in the population belongs to one of the five epidemiological groups: susceptible (\(S_i\)), exposed (\(E_i\)), infected (\(I_i\)), recovered (\(R_i\)), and dead (\(D_i\)), such that \(N_i = S_i + E_i + I_i + R_i + D_i\). 

\begin{figure}
    \centering
    \includegraphics[width=0.99\textwidth]{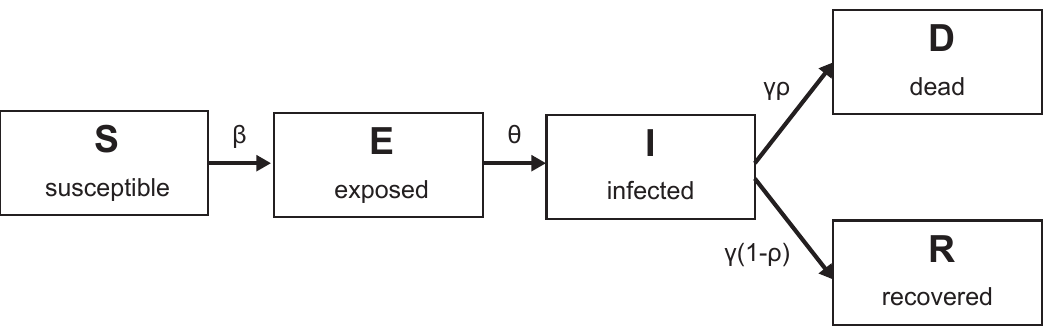}
    \caption{A schematic view of the epidemiological model which is divided into five states - susceptible, exposed, infected, recovered, and dead.}
    \label{fig:epi_model}
\end{figure}

Consumers in the susceptible group have no immunity and are susceptible to infection. When an individual in the susceptible group (\(S_i\)) is exposed to the pathogen through an interaction with an infected consumer, the susceptible consumer is transferred to the exposed epidemiological group (\(E_i\)) at a rate corresponding to the average interaction between infected consumers and susceptible consumers, denoted by \(\beta\). Each consumer stays in the exposed group on average \(\theta\) days, after which the consumer is transferred to the infected epidemiological group (\(I_i\)). Infected consumers stay in this group on average 
\(\gamma\) days, after which they are transferred to the recovered epidemiological group (\(R_i\)) or the dead (\(D_i\)) epidemiological group with probability \(\rho\)  and \(1 - \rho\), respectively. The recovered consumers are again healthy, no longer contagious, and immune from future infection. The epidemiological dynamics are formally described using a system of ordinary differential equations, as follows:

\begin{equation}
    \begin{array}{l}
    \frac{dS_i(t)}{dt} = -\beta S_i(t) I_i(t), \\ \\
    \frac{dE_i(t)}{dt} = \beta S_i(t) _iI(t) - \theta E_i(t), \\ \\ 
    \frac{dI_i(t)}{dt} = \theta E_i(t) - \gamma I_i(t), \\ \\ 
    \frac{dR_i(t)}{dt} = \rho \gamma I_i(t), \\ \\ 
    \frac{dD_i(t)}{dt} = (1 - \rho) \gamma I_i(t). \\ \\ 
    \end{array}
    \label{eq:epi_model}
\end{equation}
Importantly, \(\theta, \gamma\), and \(\rho\) are biological-clinical properties and therefore are constant between locations while \(\beta\) is highly affected by the population density, culture, and other properties making it unique for each location \cite{support_3}. Fig. \ref{fig:epi_model} presents a schematic view of the epidemiological model.

\subsubsection{Local supply and demand model}
For each location, the local supply and demand sub-model is based on the classical supply and demand model. Nonetheless, for simplicity, we assume the products' prices do not alter due to change over time. At each step in time, each consumer has a demand for products (\(\eta\)) which it can buy from firms in her location. We assume the consumer is well-informed and prefers to buy each product from the firm that provides it at the lowest price. In addition, \(\eta\) changes due to the pandemic state as a function of its sensitivity to each product as indicated by a vector \(\nu\). Formally, a consumer's overall demand at some point in time, \(t\) is \(\eta + \nu \frac{I_i}{N_i - D_i}\). Hence, at each step in time, a consumer first obtains salary, \(s\), which adds to its current available money (\(s\)) and uses it to buy products according to its current demand \(\eta + \nu \frac{I_i}{N_i - D_i}\). If some product is not available, the consumer just do not buy it. Moreover, if the consumer's available money is not enough to buy all the products according to its own demand, a random subset of products such that the consumer has enough available money to buy, is chosen. 

In addition, firms buy and sell products to other firms and sell products to consumers. Each firm is operating using consumers from the location that operates as workers. Therefore, its operational capacity is a function \(\frac{w}{w - w_i - w_d}\) where \(w_i\) and \(w_d\) are the subset of workers that are infected or dead, respectively. The operational capacity is multiplied by the duration it takes the firm to prepare a product from its ingredients. It takes the firm \(\pi\) steps in time to make produce from its ingredients (\(\upsilon\)) are available to it. At each step in time, a firm performs five actions. First, the firm pays its operational cost, \(\delta\). Second, the firm buys products from other firms, if any. Third, the firm established new supply chains, if any. Fourth, the firm prepares products from the products it acquired, if any. Finally, the firm sells its ready products to consumers. 

\subsubsection{Supply chain and spatial model}
Let us assume each location represents a single community and local economy and that products are moved between locations via supply chains, while consumers are spatially static in their communities over time. Formally, let \(G := (V, L)\) be a directional, multi-edge, and non-empty graph where \(V\) is the set of locations and \(L \subset V \times V\) is the set of supply chains. Each supply chain, \(l \in L\), is defined by two firms (\(f_1,f_2\)) located in the same or different locations and is defined by a tuple \(l := (p, d, a)\) where \(p \in P\) is the product firm \(f_1\) sell to firm \(f_2\), \(d \in \mathbb{R}^+\) is the cost to initially establish the supply chain, and \(a \in \mathbb{N}\) is the number of steps in time that takes since firm \(f_2\) buys the products and \(f_1\) provides it once these are ready for delivery. 

\subsubsection{Integrating into a single framework}
In order to numerically solve the proposed model, we implement it using the ABS approach \cite{agent_based_simulation,teddy_labib_base}. The simulation is developed using the Python \cite{python} programming language (version 3.9.2). The simulation occurred in rounds marked by \(t \in [0, T]\) such that \(T < \infty\). In the first round \((t = 0)\), the environment in the form of the \(G\) sub-model is set where locations with their initial consumer population and set of firms are generated such that the firms do not have supply chain links between them (\(|L| = 0\)). The set of products is obtained from a pre-defined distribution. In order to allow a reasonable simulation, we assume that before the pandemic, products were produced and delivered to at least satisfy the demand for each location. As such, a random subset of the firms is assumed to generate products without any ingredients (i.e., \(\upsilon = [0, \dots, 0]\)). In addition, supply chains are established at random between firms until the overall demand of the consumers is met. Finally, prices are allocated to the products without any ingredients and the selling prices of each firm are established to be the cost of all the products to generate its own with some random profit margin that allows it to end up each step in time with a profit of \(x \in [1\%,50\%]\) from its volume. 

Afterward, for each round (\(t > 0\)), the following processes occur and are performed by the agents. For the epidemiological dynamics, for each location (\(v \in V\)) the susceptible consumers become infected due to interaction with infectious consumers. Exposed consumers become infectious once \(e_t = \theta\). Infectious consumers transform to either the recovered or dead epidemiological state once \(e_t = \gamma\). Consumers that are in the dead epidemiological state (i.e., \(e = D\)) are removed from the population. In addition, for the economic dynamics, the consumers buy the products they want after obtaining a salary. Moreover, firms pay their operational cost, buy products from other firms, establish new supply chains, prepare products from the products they acquire, and sell products to consumers. 

\subsection{Supply chain resilience}
Each firm aims to make as much profit as it can (\(O_{p}\)). In our case, it means, each firm aims to fulfill the demand in all the locations it is located at over time. However, it also wishes not to bankrupt during a pandemic as long as possible (\(O_{b}\)). Intuitively, by establishing cost-optimal supply chains, a firm optimizes for the first objective while establishing supply chains with all relevant firms in the economy such that their cost is smaller than the selling price of the finished product optimizes for the latter objective. One can notice a trade-off between the two objectives. As such, a supply chain reliance strategy aims to solve the following optimization problem:
\begin{equation}
    \max_{SC} \omega_1 O_{p} + \omega_2 O_{b}
    \label{eq:task}
\end{equation}
where \(SC\) is the set of supply chains established by the firm to acquire products and \(\omega_1, \omega_2 \in [0, 1]\) are the weights of the two objectives, respectively. Formally, \(O_{p} = \frac{1}{T}\sum_{t=0}^{T} m\) and \(\min_{t} \mathbb{I}(t, m < 0)\) where \(\mathbb{I}(x, y)\) is a predict function returning \(x\) when condition \(y\) is satisfied. 

In order to solve the optimization task, we adopted the Monte Carlo approach \cite{optimization}. Namely, we sample at a random manner the combination of possible connections for each given firm. This process occurs for \(\zeta >> 1\) times, and the configuration that established the highest value for Eq. (\ref{eq:task}) is chosen. 

We focus on four supply chain strategies where no preparation for a pandemic occurs (\(\omega_2 = 0\)), where firms do not wish to make profit (\(\omega_1 = 0\)), both objectives are identically important (\(\omega_1 = \omega_2 = 0.5\)), and random case where both making profit and preparation are important to each firm in different manner (\(\omega_1 > 0, \omega_2 > 0\)). 

\subsection{Approximating supply chain strategy using machine learning}
Since the right values of \(\omega_1\) and \(\omega_2\) are strongly dependent on the economic status of the firm, its dependency on other firms in the economy, the consumers' behavior, and the magnitude of the pandemic, it is only reasonable that different firms will adopt different \(\omega_1, \omega_2\) values as part of their supply chain reliance strategy. Nonetheless, solving such a question analytically is unrealistic as one would be required to solve for the entire economy at once and would be highly sensitive to any change.  

Thus, in order to find the (near) optimal  \(\omega_1, \omega_2\) values for each firm, we adopted a data-driven approach. Namely, using experimental data, one can use a machine learning (ML) based model to fairly approximate the \(\omega_1, \omega_2\) values of a firm without actually solving for the exact scenario due to the generalization capabilities of ML models \cite{ml_generalization_1,ml_generalization_2,ml_generalization_3}. Formally, this is a one-dimentional regression problem since  \(\omega_1 + \omega_2 = 1\) which infers that by finding \(\omega_1\), one can directly compute \(\omega_2\). In order to use a ML model, one is first required to collect representative data. To this end, we run the simulation multiple times with different parameter values. For each such run, we use the MC approach for the values of \(\omega_1, \omega_2\) for each firm. As a target variable for the ML to predict, we use \(O_p + O_b\), ignoring the values of \(\omega_1, \omega_2\) in the objective to obtain a consistent evaluation of the firm's performance over different runs. 

Using the obtained dataset, we use the Tree-based Pipeline Optimization Tool (TPOT) \cite{tpot}, an automated machine learning tool that uses genetic algorithms \cite{ga_intro} to optimize ML models. TPOT tries multiple ML models on the dataset to find the one that performs best. In order to make sure the results are robust, we adopted the k-fold cross-validation method with \(k=5\) \cite{pick_k}. Since the obtained model is black-box \cite{deep_good}, we used two methods to explore how the model allocates \(\omega_1\) values to firms. First, we use the information gain feature importance method \cite{information_gain} to learn how much each feature of the economy influences the model's prediction. Second, SHapley Additive exPlanations (SHAP) analysis was used to gain insight into the influence of various features \cite{clinical_shap}. The SHAP values can be used to explain the output of a ML model by attributing the contribution of each individual feature to a particular prediction \cite{shap_value}. 

\subsection{Experimental setup}
Due to the difficulty of obtaining realistic data on firms' supply chain and financial decisions, one can explore a large number of synthetic economies to obtain statistically representing dynamics. Thus, we explore the influence of a pandemic in different levels of magnitude on an arbitrary economy. For simplicity, for each sample configuration of an economy, we run the ABS simulation for \(n=100\) time to obtain a statistically represented sample. For the epidemiological-related parameter values, we used values associated with the COVID-19 pandemic \cite{covid_19_exmple_data_teddy}. Table \ref{table:params} shows the parameter values used in the experiments. Importantly, we assume all days are working days (i.e., not considering weekends and holidays). 

\begin{table}[!ht]
\centering
\begin{tabular}{lll}
\hline \hline
\textbf{Parameter} & \textbf{Symbol} & \textbf{Value range} \\ \hline \hline
simulation duration & \(T\) & 365 \\
Duration of a step in time & \(\Delta t\) & 1 day \\
Monte Carlo repetition count  & \(\zeta\) & 1000 \\
Number of consumers in a location & \(|N_i|\) & 500-5000 \\
Number of firms in a location & \(|F_i|\) & 5-50 \\
Number of locations & \(|V|\) & 1-20 \\
Initial amount of money & \(m \; (t=0)\) & \(1 \cdot 10^2 - 5  \cdot 10^6\) \$ \\
Salary & \(s\) & \(5.5 \cdot 10^1 - 5.5 \cdot 10^3\) \$ \\
Number of unique products in the economy & \(|P|\) & 10-250 \\
Firms initial available money & \(\varphi; (t=0)\) & \(1 \cdot 10^4 - 5 \cdot 10^7\) \$ \\
Firms operational cost & \(\delta\) & \(0.0025\varphi - 0.025\varphi\) \$ \\
Workers in a firm & \(w\) & \(1 - 1000\) \\
Price of a product & \(\iota\) & \(1 \cdot 10^{0} - 1 \cdot 10^4\) \$ \\
Location's average infection rate & \(\beta\) & \(5 \cdot 10^{-5} - 1 \cdot 10^{-2}\)  \\
During from exposed to infectious & \(\theta\) & \(5-9\) days  \\
During from infection to recovered/dead & \(\gamma\) & \(10-18\) days  \\
Recovery rate & \(\rho\) & \(0.975-0.995\) \$ \\
Ingredients per product & \(\) & 1-20 \\
Number of products a firm sell to consumers & \(\) & 1-10 \\
TPOT population size & \(\) & 50 \\
TPOT number of generations & \(\) & 20 \\
Number of simulations for the machine learning model & \(\) & 500 \\
Number of \(\omega_1, \omega_2\) configurations for each machine learning sample & \(\) & 20 \\
 \hline \hline
\end{tabular}
\caption{The model's parameter value ranges used in the experiment.}
\label{table:params}
\end{table}

\section{Results}
\label{sec:results}
In this section, we present the results of the experiments. First, we show the pandemic effect on firms over time for four supply chain resilience strategies. Second, we show a sensitivity to important pandemic-related parameters. Finally, we show the obtained ML model to estimate \(\omega_1\) for firms.  

\subsection{Pandemic effect of supply chain over time}
\begin{figure}
    \centering
    \includegraphics[width=0.99\textwidth]{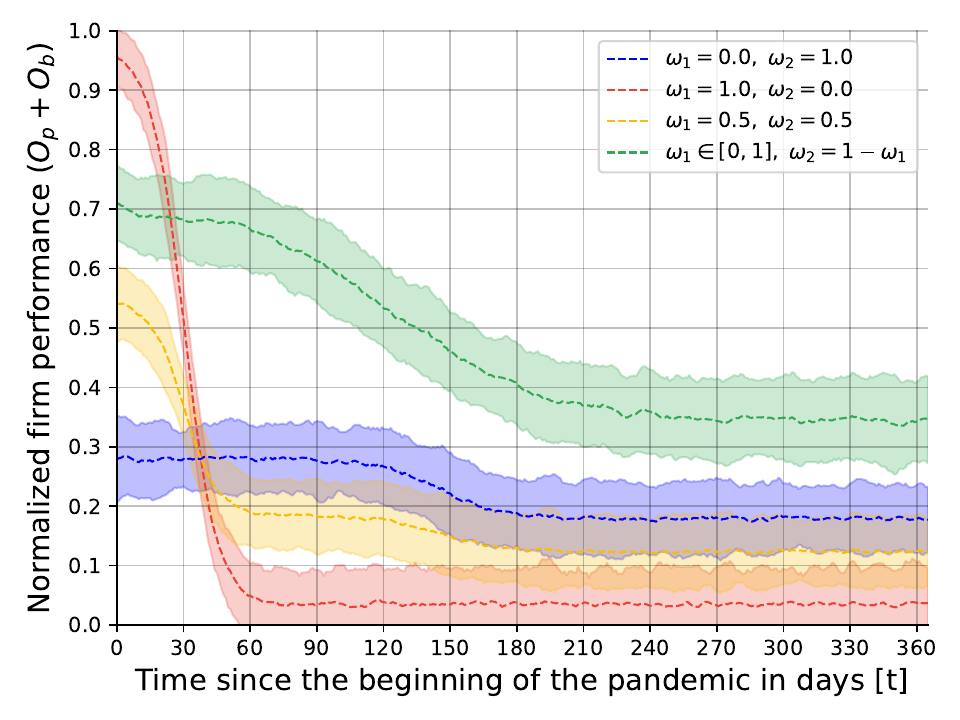}
    \caption{Analysis of four supply chain resilience strategies as the course of a one-year-long pandemic. The results are shown as the mean \(\pm\) standard deviation of \(n=1000\) simulations. }
    \label{fig:first_result}
\end{figure}

We start by investigating the four configurations of supply chain resilience over time during a pandemic. Fig. \ref{fig:first_result}.prsents this analysis where the x-axis is the time in days that passed since the beginning of the pandemic and the y-axis is the normalized firm performance as outlined in Eq. (\ref{eq:task}) where \(\omega_1\) and \(\omega_2\) are agnostic to make allow the comparison between the four different strategies. One can notice that firms that only consider the preparation for the pandemic (\(\omega_1 = 0\)) are very inefficient overall as these start around 0.3 while less affected by the pandemic as after a year the average performance is around 0.2. Unalike, when firms do not prepare their supply chains for a pandemic at all (\(\omega_1 = 1\)), the firm's performance is near optimal but after only two months of the pandemic the average performance drops to around 0.04 which is almost full economic collapse. The strategy that all firms balance the two strategies results in an average performance between the two previous cases where the initial performance is around 0.55 and after 40 days drops to around 0.2 performance followed by a further slower decline towards 0.1 after around 200 days. For the heterogeneous case where each firm aims to find its balance of \(\omega_1\) and \(\omega_2\), the initial performance is the second highest with a score around 0.7. During the time of the pandemic, the performance decreases relatively slowly and balanced after around 240 days around 0.35 - the highest performance of the four strategies. 

\subsection{Pandemic-related sensitivity analysis for supply chain resilience}
\begin{figure}[!ht]
    \begin{subfigure}{.5\textwidth}
        \includegraphics[width=0.99\textwidth]{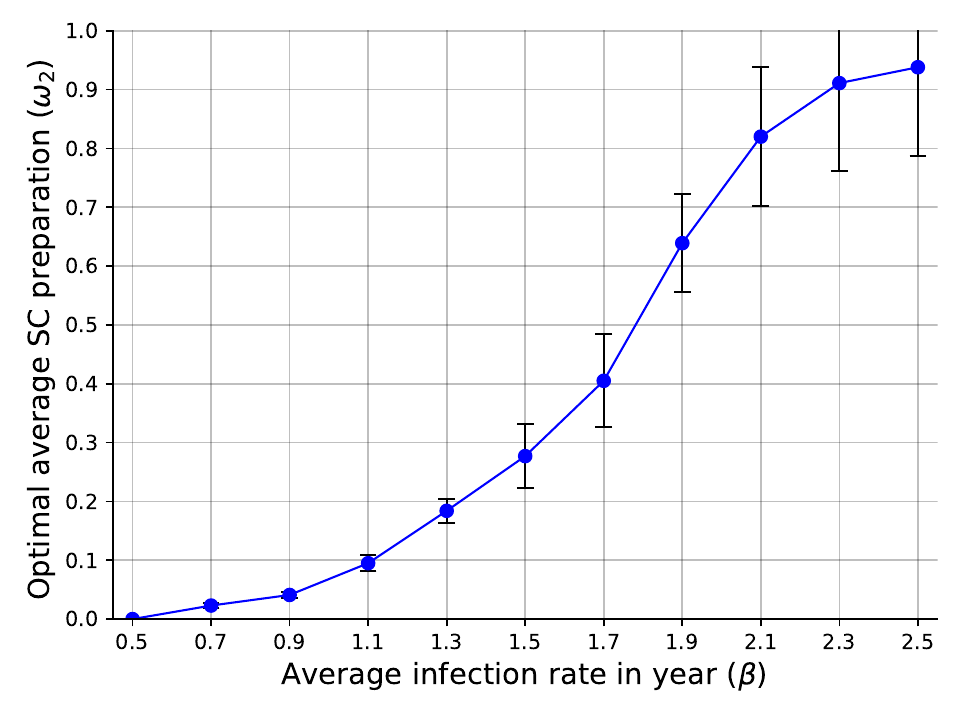}
        \caption{Average infection rate in year (\(\beta\)).}
        \label{fig:sens_beta}
    \end{subfigure}
    \begin{subfigure}{.5\textwidth}
        \includegraphics[width=0.99\textwidth]{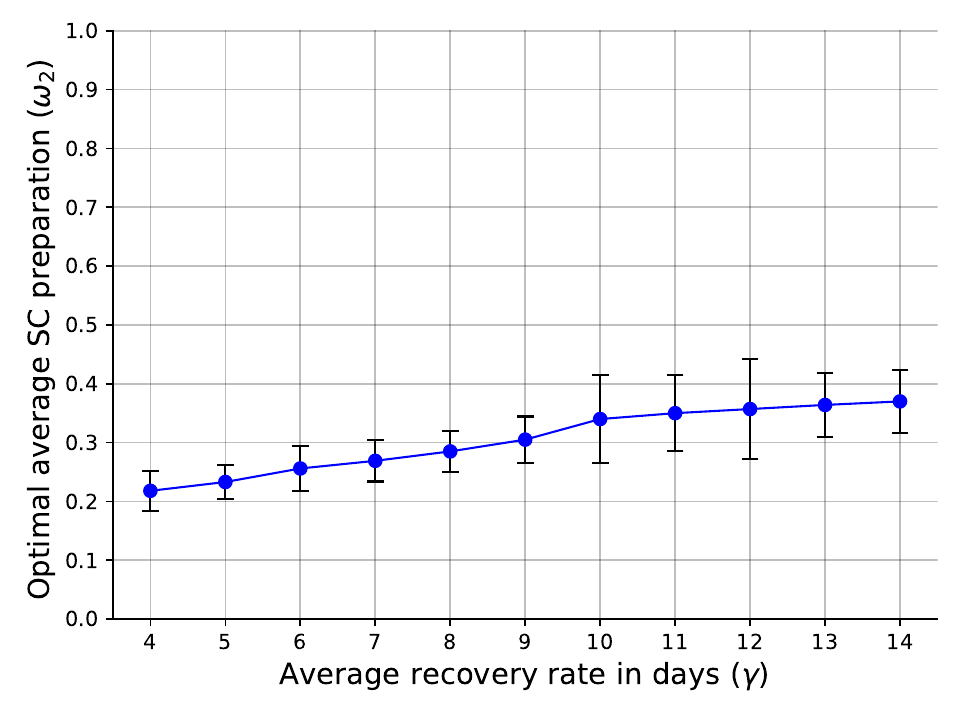}
        \caption{Average recovery rate in days \(\gamma\).}
        \label{fig:sens_gamma}
    \end{subfigure}
    
    \begin{subfigure}{.5\textwidth}
        \includegraphics[width=0.99\textwidth]{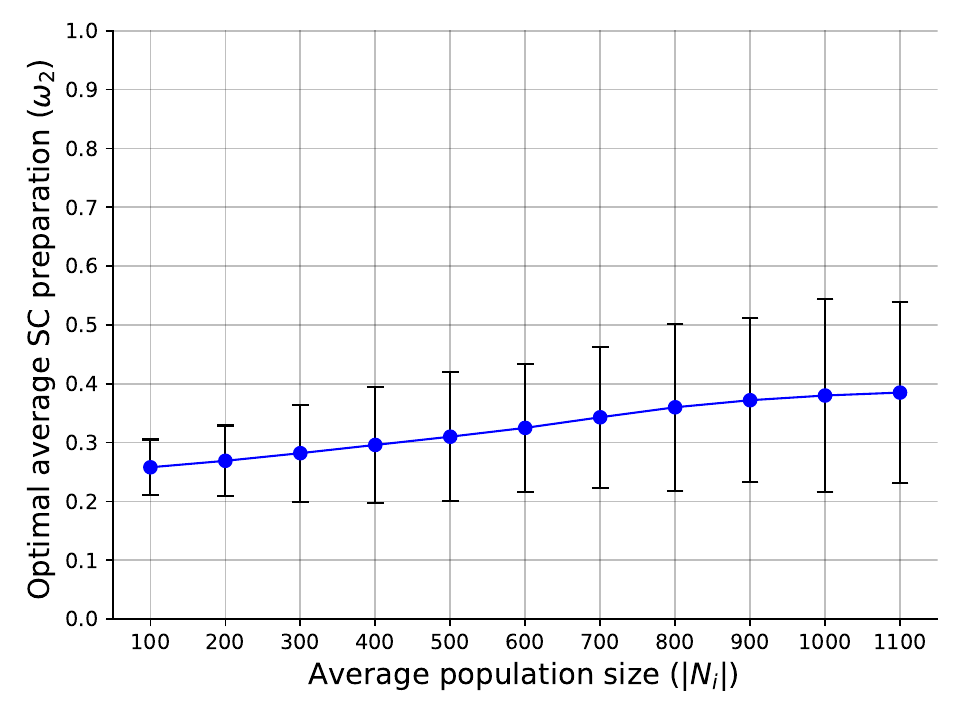}
        \caption{Average population size in a location (\(|N_i|\)).}
        \label{fig:sens_n}
    \end{subfigure}
    \begin{subfigure}{.5\textwidth}
        \includegraphics[width=0.99\textwidth]{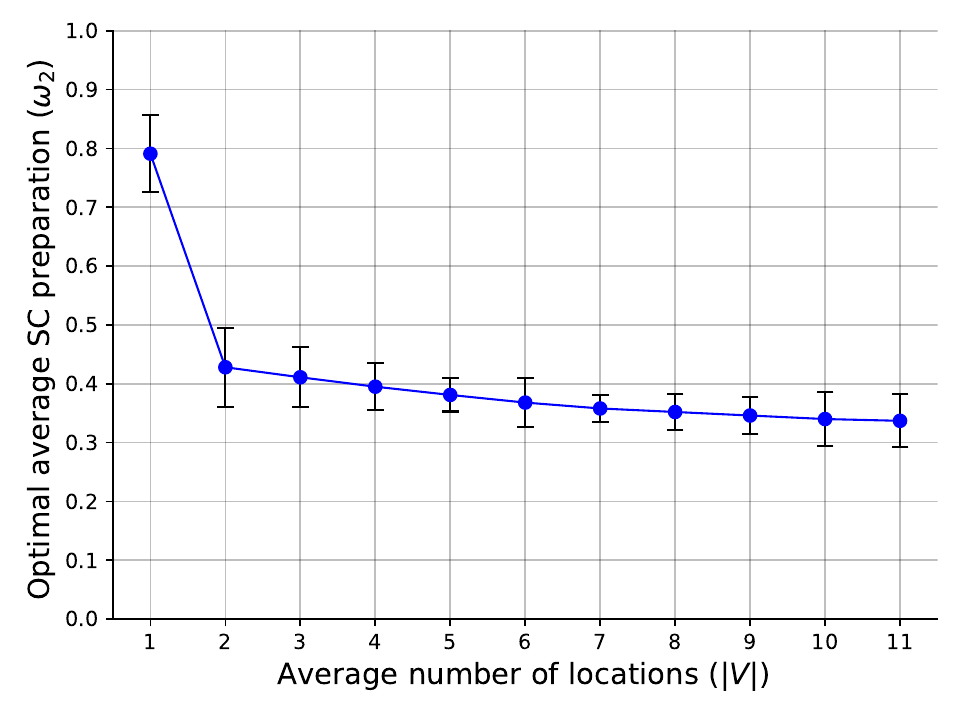}
        \caption{Number of locations (\(|V|\)).}
        \label{fig:sens_locations}
    \end{subfigure}
    
    \caption{A sensitivity analysis of the model's main epidemiologically-related parameters. The weight of profit rather than supply chain resilience parameter (\(\omega_1\)) is obtained as an average \(\pm\) standard deviation of \(n=100\) simulations with the rest of the parameters are sampled according to Table \ref{table:params}.}
    \label{fig:sens}
\end{figure}

Let us focus on the most realistic case of the four strategies out of the four - where each firm has its own values for \(\omega_1\) and \(\omega_2\). Since finding the optimal value for \(\omega_1\) is extremely computationally challenging, compute a sensitivity analysis such that each case shows \(n=100\) unique runs, taking the average best \(\omega_1\) value. Fig. \ref{fig:sens} shows the results of the sensitivity analysis such that the x-axis indicates the parameter under investigation and the y-axis indicates the optimal average \(\omega_1\) value. The results are shown as the mean \(\pm\) standard deviation of the \(n=100\) runs for each parameter value. Specifically, Fig. \ref{fig:sens_beta} shows the change of the optimal average \(\omega_1\) as a function of the average infection rate (\(\beta\)). One can notice a sigmoid function with the standard deviation growing alongside the value of \(\omega_1\). Fig. \ref{fig:sens_gamma} focus on the change due to the average recovery rate (\(\gamma\)) where a higher \(\gamma\) value results in higher \(\omega_1\) values in a somewhat linear fashion. Similarly, Fig. \ref{fig:sens_n} shows that \(\omega_1\) is linearly increasing with the increase in the average population size but this outcome is increasingly less accurate as the population grows which is indicated by the increasing in the error bars' size. Fig. \ref{fig:sens_locations} reveals a sharp decrease in the value of \(\omega_1\) between one and two locations while afterward the value of \(\omega_1\) decreases logarithmically with respect to the value of the average number of locations. 

\subsection{Supply chain resilience machine learning model}
\begin{figure}
    \centering
    \includegraphics[width=0.99\textwidth]{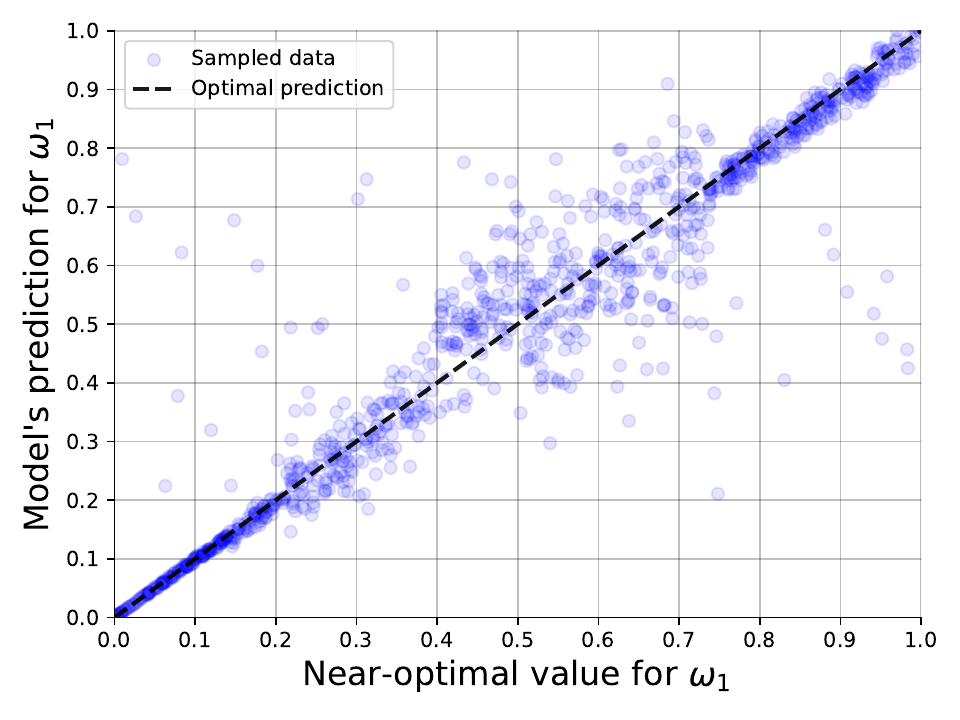}
    \caption{The machine learning model's performance for 1000 randomly picked predictions for near-optimal value of \(\omega_1\) as obtained by the Monte Carlo method with the agent-based simulation compared to the obtained machine learning model's prediction. The dashed line indicates the locations of perfect predictions.}
    \label{fig:ml_perofrmance}
\end{figure}

Overall, we trained the ML model on 2.2 million samples of companies \(\omega_1\) value, \(O_p + O_b\), and the economy's initial state before the beginning of a pandemic. An additional 0.55 million samples are used to evaluate the obtained model's performance. The error analysis of the model shows that the obtained model has a mean absolute error of \(0.047\) and a coefficient of determination of \(R^2 = 0.816\). Fig. \ref{fig:ml_perofrmance} presents the ML model's prediction for 1000 randomly picked samples. The x-axis indicates the near-optimal value for \(\omega_1\) as obtained using the MC process using the ABS simulation while the y-axis is the ML model's prediction for \(\omega_1\) for the same firm given the same initial conditions used by the ABS simulation. One can notice that for \(\omega_1 < 0.2\) and \(\omega_1 > 0.7\) the ML model makes very accurate predictions while for \(0.2 < \omega_1 < 0.7\) the model has a larger error, on average. In addition, some firms have a relatively large error but these are a relatively small portion of the entire dataset. 

\begin{figure}
    \centering
    \includegraphics[width=0.99\textwidth]{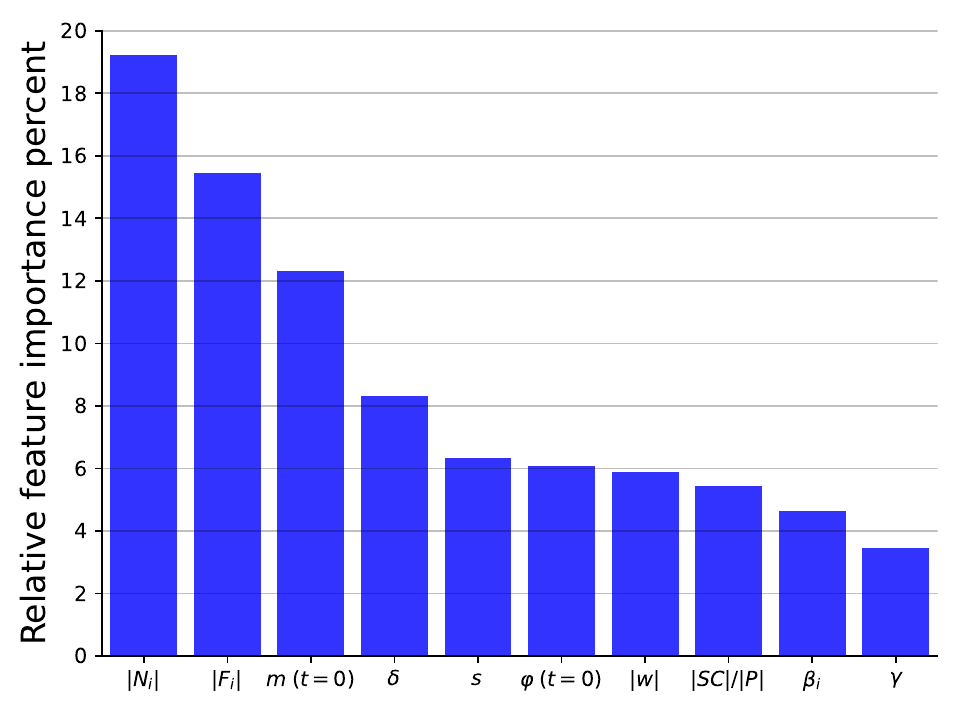}
    \caption{Feature importance analysis of the obtained machine learning model. The results show the average value of a 5-fold cross-validation analysis. These features are responsible for 87.13\% of the model's decision making process.}
    \label{fig:ml_fi}
\end{figure}

Fig. \ref{fig:ml_fi} shows the feature importance distributions of the top 10 most important features which are responsible for 87.13\% of the properties the ML uses to make a prediction. The most important feature with 19.23\% is the number of consumers in a location the firm operates in, followed by the number of firms in the location (15.44\%). Afterward, the initial available money of the consumers and the initial available money of the company are the third and sixth most important features with 12.32\% and 6.08\%, respectively. The operational cost and consumer's average salary are the forth and fifth most important features with 8.31\% and 6.34\%, respectively. The pandemic-related parameters, the average infection rate (\(\beta_i\)), and recovery rate (\(\gamma\)) are the ninth and tenth most important features with 4.65\% and 3.44\%, respectively. The operational features - the number of workers and the number of supply chains divided by the number of products in the economy are the seventh and eights most important features with 5.87\% and 5.45\%, respectively. 

\begin{figure}
    \centering
    \includegraphics[width=0.99\textwidth]{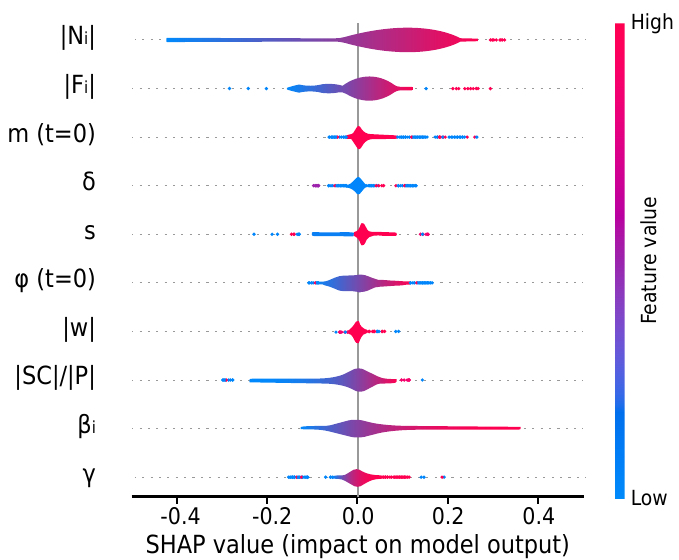}
    \caption{A Shap analysis of the ML model. Dots on the right side indicate a contribution towards the model predictions of \(\omega_1 = 0\) while left side dots indicate a contribution towards the model predictions of \(\omega_1 = 0\). }
    \label{fig:ml_shap}
\end{figure}

Fig. \ref{fig:ml_shap} shows the Shap analysis of the obtained ML model. On the x-axis, dots on the right side indicate a contribution towards the model predictions of \(\omega_1 = 0\) while left side dots indicate a contribution towards the model predictions of \(\omega_1 = 0\). One can notice a mostly consistent behavior for \(|N_i|\) and \(|F_i|\) while the other feature demonstrate a more chaotic pattern. 

\section{Discussion and Conclusion}
\label{sec:conclusion}
In this study, we proposed a novel agent-based simulation (ABS) based model for the influence of pandemic spread on supply chain resilience strategies. The spatio-temporal model combined an extended SIR epidemiological model (SEIRD) integrated with a supply-demand economic model for multiple physical locations. This model is used to evaluate the efficiency of different supply chain resilience on a wide range of scenarios using \textit{in silico} experiments, focusing on balancing between profit maximization and preparation for a pandemic in terms of supply chain resilience strategy. 

Using this model, we started by exploring the \say{edge} cases of supply chain resilience as well as the average and heterogeneous cases to evaluate the possible range of strategies, as shown in Fig. \ref{fig:first_result}. Unsparingly, for the case where firms adopted a supply chain resilience strategy of doing nothing, a pandemic results in near to the collapse of the entire economy while the other end of the rope results in the least effect by the pandemic but provides a strongly unappealing and inefficient economic environment. A more balanced case performs better for times without a pandemic but is still almost half as good as the first strategy while in the end decreases to only slightly than twice as good performance from the first case just after half a year. A heterogeneous case provides both a more realistic and better-performing strategy for firms in an economy. This outcome, while unsurprising, empirically supports the further analysis of this supply chain resilience strategy and aligns to previous works that highlighted the benefits of heterogeneous decision-making of agents to solve complex tasks \cite{dis_first_result_1,dis_first_result_2}.

When considering the most realistic case of these where each firm has a unique strategy of supply chain resilience for pandemics, the model predicts somewhat expected results, as revealed by Fig. \ref{fig:sens}. Initially, as the infection rate growth, firms are required to be more prepared to a pandemic to overcome it. This result is intuitive and very well established \cite{airborne_sample_1,airborne_sample_2}. In a similar manner, increased in the recovery rate requires more preparation as the overall number of infected individuals is higher at the same time which further contributes to the pandemic spread \cite{brn}. Population size increases cause more drastic shifts in an economy for the same pandemic, on average, which requires more supply chain resilience and also generates more uncertainty as more infection paths can occur \cite{density_bad}. Interestingly, one can observe a sharp decrease in the supply chain resilience between one and two locations, as shown in Fig. \ref{fig:sens_locations}. This drop in the value of \(\omega_2\) can be associated with the fact that not the entire pandemic occurs at once and influences the entire economy and therefore, a less dramatic supply chain resilience strategy can be adopted by firms to overcome the same pandemic, on average. This effect is greatly reduced between two and more locations as visible in the same plot. 

We aimed to provide an approximation tool for firms to manage their supply chain resilience strategy with respect to their state and the economy's state using an ML model. As illustrated in Fig. \ref{fig:ml_perofrmance}, the obtained ML model is extremely accurate for firms that are not significantly affected by a pandemic (of a small-medium size), and its performance is reduced as the firm is more sensitive to the pandemic. That said, for very sensitive firms, the ML model is again performing relatively well. This outcome can be explained by the fact that for the more extreme cases, it is clearer to ignore pandemic risks or be extremely prepared for them while the intermediate cases are more chaotic in nature, resulting in a wide range of possible outcomes \cite{sc_disucssion_ml_1}. Interestingly, the ML model predicts, as shown in Fig. \ref{fig:ml_fi}, that the economy size, as reflected by the number of consumers and firms, is the most important parameter to the individual firm strategy - agreeing with previous studies about supply chain resilience \cite{sc_disucssion_ml_3,sc_disucssion_ml_4}. The firm's size and economic state, as indicated by the number of workers and the operational cost, are very important but less than the overall economy's dynamics as also found by \cite{sc_disucssion_ml_2}. However, this average importance value hides a more chaotic picture which is revealed by Fig. \ref{fig:ml_shap}. While the economic size has mostly a positive effect on the number of how much a firm should prepare for a pandemic, the rest of the features show inconsistent behavior when considered individually. This outcome can be expected due to the complex connections between the features in the dynamics that dictate a firm's future and therefore optimal strategy before a pandemic starts. Nonetheless, it highlights the importance of other studies to explore multiple features for supply chain resilience at once rather than one at a time.

The proposed model is not without limitations. The proposed model assumes a pandemic originated with a single pathogen that does not mutate over time. While this assumption is commonly used \cite{multi_strain_1,multi_strain_2,multi_strain_4,multi_strain_3} it is known to be false for even medium-size pandemics, and future works should use multi-strain with re-infection mechanism epidemiological sub-model to obtain more realistic epidemiological dynamics. Second, we assume that the prices of products from the firms are static over time to focus on the pandemic spread influence on the supply chain rather than other factors. Nonetheless, natural price fluctuations as well as high-order pandemic-related price fluctuations may play a central role in supply chain management and should be included in future work to make it more realistic. Third, population size and the number of firms as well as their spatial distribution are assumed to be static over time. Adding consumer migration dynamics as well as the introduction of firm establishment and closer are also promising venues for future work. Fourth, the pandemic-related demand assumes all consumers are aware of the pandemic state in their community at any given point in time. A relaxation of this assumption by adding delay and only an estimation for the pandemic spread would make the model more realistic and reflect the actual available information consumers have during a pandemic. 

Taken jointly, the proposed model and its agent-based simulator provide policymakers and business owners with a computational tool to evaluate their supply chain preparedness for the event of a large-scale pandemic, such as COVID-19. Our results show that for even relatively large pandemics, well-prepared businesses are theoretically able to overcome the challenge and thrive during and after the pandemic ends.

\section*{Declarations}
\subsection*{Funding}
This research did not receive any specific grant from funding agencies in the public, commercial, or not-for-profit sectors.

\subsection*{Competing interests}
The authors have no relevant financial or non-financial interests to disclose.

\subsection*{Acknowledgment}
The author wishes to thank Labib Shami for motivating this research and providing valuable economic advice. 

\bibliographystyle{elsarticle-num-names} 
\bibliography{cas-refs}

\end{document}